\RequirePackage{fix-cm}
\documentclass[smallextended]{svjour3}       
\smartqed  
\usepackage[dvipdfm,  
            pdfstartview=FitH,
            CJKbookmarks=true,
            bookmarksnumbered=true,
            bookmarksopen=true,
           colorlinks, 
            pdfborder=001,   
            linkcolor=blue,
            anchorcolor=blue,
            citecolor=blue
            ]{hyperref}
\usepackage{lscape}
\usepackage{amsmath}
\usepackage{bm}

\usepackage{graphicx}
\usepackage{array}
\usepackage{booktabs}
\usepackage{subfigure}
\usepackage{caption}
\usepackage{fancyhdr}
\usepackage{array}
\usepackage{geometry}
\usepackage{extarrows}
\usepackage{multirow}
\usepackage{listings}
\begin{document}

\newpage
\setcounter{page}{1}
\title{Generating correlated random vector by polynomial normal transformation
}

\author{Qing Xiao   
}

\institute{Qing Xiao \at
              School of Mechatronic Engineering and Automation, Shanghai University, Shanghai, 200072, China \\
              \email{xaoshaoying@shu.edu.cn}           
}

\date{Received: date / Accepted: date}

\maketitle

\begin{abstract}
This paper develops a polynomial normal transformation model, whereby various non-normal probability distributions can be simulated by the standard normal distribution. Two methods are presented to determine the coefficients of polynomial model: (1) probability weighted moment (PWM) matching (2) percentile matching. Compared to the existing raw moment or L-moment matching, the proposed methods are more computationally convenient, and can be used to estimate the coefficients of polynomial model with a higher degree. Furthermore, for two correlated random variables, a polynomial equation is derived to estimate the equivalent correlation coefficient in standard normal space,  and random vector with non-normal marginal distributions and prescribed correlation matrix can be generated. Finally, numerical examples are worked to demonstrate the proposed method.
\end{abstract}

\section{Introduction}
In order to generate non-normal random variables, many families of distributions have been proposed. \cite{NonormalCompare} summarizes the related works, and compares six procedures in terms of the efficiency, simplicity and generality. Among these approaches, two standard normal distribution based transformation models: Johnson system and third-order polynomial normal transformation (TPNT) technique, demonstrate the potential utility for generating correlated random variables.

\cite{Johnson} discusses the feasibility of simulating continuous non-normal distributions by standard normal distribution, and develops a four-parameter transformation system. The Johnson system comprises four transformation models: $S_N$ (Normal distribution), $S_L$ (Lognormal family), $S_U$ (Unbounded  family) and $S_B$ (Bounded  family), which can be employed to approximate many non-normal distributions. For the detailed introduction to Johnson system, one can refer to literature\cite{Johnson2}.

Johnson system has been widely used to  generate non-normal random numbers, but it shows inadequacy of accommodating the dependence among random variables. If samples of the  correlated random vector are available, the Johnson system is workable. By normalizing each sample value through the corresponding transformation model, the correlation matrix in the normal space can be calculated directly ( see Section 3.2 of reference\cite{Johnson3}. However, this approach cannot be used to generate random vector with assigned correlation matrix. To achieve this goal, it is necessary to analytically estimate the correlation coefficient $\rho_z$ in normal space for a given correlation coefficient $\rho_x$ between two random variables.

Compared with Johnson system, the dependence among random variables can be easily handled by TPNT technique. For a given $\rho_x$, the analytical equation can be derived to estimate $\rho_z$, and random vector with desired correlation matrix can be conveniently generated\cite{TPNTC1,TPNTC2H}. The TPNT is firstly introduced by \cite{TPNT}, this technique generates non-normal deviates by a polynomial transformation of standard normal ones. A detailed investigation of TPNT is given in reference\cite{TPNTHongKong}, it is found there are many probability distributions that TPNT technique cannot simulate.

To expand the generality of the polynomial transformation technique, a fifth-order polynomial normal transformation (FPNT) technique is developed in reference \cite{FPNT}. Due to the ability to control the extra fifth and sixth order moments,  FPNT can approximate some distributions that are difficult for TPNT\cite{FPNT_PDF}, but it requires tedious mathematical computation to determine the  coefficients of FPNT. By equating the first six raw moments of FPNT model with those of the modeled distribution, a system of nonlinear equations is established. The coefficients are estimated by solving the equations numerically. In pursuit of a simpler parameter estimation method, L-moments are employed to characterize FPNT model\cite{FNPTLmoments}, and the analytical expression of coefficients are obtained. However, because of the tedious mathematical derivation, this L-moments based method is unable to estimate the coefficients of a polynomial model with a higher degree, which may be used to simulate more non-normal distributions.

In this paper, the polynomial normal transformation model is extended to a much higher degree. Two methods: probability weighted moment (PWM) matching and percentile matching, are developed to assess the coefficients of polynomial model. By equating PWMs of the polynomial model with those of the modeled distribution, a system of linear equations solved for the coefficients is established. For the percentile matching, the polynomial model can be established by a least square method, and a numerical investigation is also performed to check the generality and  accuracy of this method.

Furthermore, using the product moments formulae of two correlated standard normal variables \cite{Pearson}, the polynomial equation of $\rho_z$ is derived. For a given $\rho_x$ between two correlated random variables, an appropriate value of $\rho_z$ is conveniently determined, and random vector with assigned correlation matrix can be generated from standard normal deviates. Through numerical examples, it makes evident that non-normal distributions can be well simulated by the polynomial model, and random vector with prescribed correlation structure can be generated  by the proposed procedures.

\section{Polynomial normal transformation model}
Let $X$ be a continuous random variable with cumulative distribution function (CDF) $F(X)$, let $Z$ be a standard normal random variable with CDF $\Phi(Z)$. Equate the two CDFs:
\begin{equation}
  F(X)=\Phi(Z)
\end{equation}
Then, $X$ is expressed as:
\begin{equation}\label{Nataf}
  X=F^{-1}\left[\Phi(Z)\right]
\end{equation} 	
where $F^{-1}(\cdot)$ denotes the inverse CDF of $X$.

Approximate the function $F^{-1}\left[\Phi(Z)\right]$ by a polynomial of degree $n$:
\begin{equation}\label{Polynomial}
 X\simeq \sum_{k=0}^{n}a_kZ^k
\end{equation}
where $a_k$ ($k=0,\dots,n$) are the undetermined coefficients. Setting $n=3$ or $5$ would yield the TPNT or FPNT model.

Since  both $F^{-1}(\cdot)$ and $\Phi(Z)$ are monotonic increasing functions, so is the composite function $F^{-1}\left[\Phi(Z)\right]$. According to Weierstrass approximation theorem\cite{saxe2002beginning}, the monotonic smooth function $F^{-1}\left[\Phi(Z)\right]$ can be well approximated by a polynomial of $Z$ over a  closed interval. While, the major problem is to determine a set of coefficients $a_k$ ($k=0,\dots,n$), such that the random variable $X$ would be well simulated by the standard normal variable $Z$. Most often, the researchers resort to the  raw moment matching method\cite{TPNT,FPNT}, which requires tedious mathematical computation. For a polynomial of degree 5, the resulted system of nonlinear equations is very complicated (see Appendix A of reference\cite{FPNT}), let alone for a polynomial model of higher degree. In this paper, the PWM matching and percentile matching are introduced to simplify the coefficients estimation problem, and the polynomial model is extended to higher degree, allowing for the simulation of more probability distributions.
\subsection{Probability weighted moment matching method}

The PWM of $X$ is defined as \cite{PWM1}:
\begin{equation}
  M_{r,p,s}=E\left\{ \left[F(X)\right]^r\cdot X^p\cdot \left[1-F(X)\right]^s\right\}
\end{equation}
For computational convenience, a particular type of PWM, $\beta_r=M_{r,1,0}$, is considered:
\begin{equation}\label{PWM1}
  \beta_r=E\left\{ \left[F(X)\right]^r\cdot X \right\}=\int_{-\infty }^{+\infty } F^r(X)\cdot X\cdot f(X)dX
\end{equation}
where $f(X)$ is the probability density function (PDF).

If the PDF of $X$ is unknown, only a set of data is available. Sort the sample into ascending order: $x_1 \leq \dots \leq x_i \leq \dots \leq x_m$, the unbiased estimate of $\beta_r$ is \cite{PWM2}:
\begin{equation}\label{PWM2}
 \beta_r=\frac{1}{m}\sum_{i=r+1}^{m}\frac{(i-1)(i-2)\cdots (i-r)}{(m-1)(m-2)\cdots (m-r)}x_i
\end{equation}

Substitute Eq.\eqref{Polynomial} into Eq.\eqref{PWM1}, the PWM of polynomial model is:
\begin{equation}
  \begin{split}
    \beta_r &=\int_{-\infty}^{+\infty}\Phi^r(Z)\cdot\left(\sum_{k=0}^{n}a_kZ^k\right)\cdot  \varphi(Z)dZ \\
      & =\sum_{k=0}^{n}a_k\int_{-\infty}^{+\infty}\Phi^r(Z)\cdot Z^k\cdot  \varphi(Z)dZ\\
      & =\sum_{k=0}^{n}a_kM^{z}_{r,k,0}\\
      M^{z}_{r,k,0}&=\int_{-\infty}^{+\infty}\Phi^r(Z)\cdot Z^k\cdot \varphi(Z)dZ
  \end{split}
\end{equation}
where $\varphi(Z)$ is the PDF. Since $\Phi(Z)$ and $\varphi(Z)$ are known functions,  $M^{z}_{r,k,0}$ can be integrated numerically.

For an $n$th order polynomial model, perform  the first $(n+1)$ PWMs matching, the following system of linear equations is established:
\begin{equation}\label{PWMequation}
\begin{pmatrix} M^{z}_{0,0,0}    & \cdots  & M^{z}_{0,k,0}  & \cdots & M^{z}_{0,n,0} \\
                \vdots           & \cdots  & \vdots         & \cdots & \vdots        \\
                M^{z}_{r,0,0}    & \cdots  & M^{z}_{r,k,0}  & \cdots & M^{z}_{r,n,0} \\
                \vdots           & \cdots  & \vdots         & \cdots & \vdots        \\
                M^{z}_{n,0,0}    & \cdots  & M^{z}_{n,k,0}  & \cdots & M^{z}_{n,n,0}
  \end{pmatrix}
 \cdot \begin{pmatrix} a_0 \\
                \vdots    \\
                a_k \\
                \vdots      \\
                a_n
  \end{pmatrix} =\begin{pmatrix} \beta_0 \\
                \vdots    \\
                \beta_r \\
                \vdots      \\
                \beta_n
  \end{pmatrix}
\end{equation}
Calculate the first $(n+1)$ PWMs of $X$:
 \begin{equation}\label{PWM2222}
  \beta_r=\int_{-\infty }^{+\infty }F^r(X)\cdot X\cdot f(X)dX=\int_{0}^{1}F^{-1}(p)\cdot p^rdp
\end{equation}
Substitute these PWMs into Eq.\eqref{PWMequation}, solving the system of linear equations gives the coefficients $a_k$.

Intuitively, a polynomial of higher degree allows for the control of higher order moments, and should give a better approximation of the target distribution. However, as the degree of the polynomial $n$ increases,  the coefficient matrix becomes near singular. Below is the determinant of the coefficient matrix:
\setcounter{table}{0}
\begin{table}[hptb]
\centering
\caption{The determinant of  coefficient matrix}
\begin{tabular}{cccc}
\hline $n$  & Determinant &$n$  & Determinant   \\
\hline
$1$  & $0.2821$                & $7$      &$3.3321\times 10^{-14}$   \\
$2$  & $0.0259$                & $8$      &$7.1266\times 10^{-18}$   \\
$3$  & $8.2291\times 10^{-4}$  & $9$      &$5.7686\times 10^{-22}$    \\
$4$  & $9.3220\times 10^{-6}$  & $10$     &$1.7762\times 10^{-26}$    \\
$5$  & $3.8458\times 10^{-8}$  & $11$     &$2.0880\times 10^{-31}$    \\
$6$  & $5.8600\times 10^{-11}$ & $12$     &$9.4023\times 10^{-37}$    \\
\hline
\end{tabular}
\end{table}
\newline For a system of linear equations with a near singular coefficient matrix, the solution is very sensitive to small errors in the estimation of $M^{z}_{r,k,0}$ or $\beta_r$. Thus, when modeling theoretical distributions,  a polynomial model of degree higher than 12 is not recommended, otherwise, the near singular coefficient matrix would yield biased values of $a_k$, giving rise to an undesirable simulation. Similarly, when used for data fitting, a  polynomial of lower degree is more preferable, in case of the error caused by the high-order PWMs drawn from the samples (In \cite{zou2014solving}, a 9th-order polynomial model is employed).

\subsection{Percentile matching method}
For a polynomial model, the following requirement should be satisfied:
\begin{equation}
  x_p=a_0+a_1z_p+\cdots+a_nz_p^n
\end{equation}
where $p$ is a percentage ($p\in[0,1]$), $x_p$ and $z_p$ are the corresponding percentiles from the target distribution $F(X)$ and the standard normal distribution $\Phi(Z)$:
\begin{equation}\label{Inverse}
  x_p=F^{-1}(p)~~~~~~z_p=\Phi^{-1}(p)
\end{equation}
Select $m$ values of $p_i$, $m\geq (n+1)$, calculate the associated percentiles: $z_{p_i}$ and $x_{p_i}$ ($i=1,\cdots,m$). Based on $m$ pair values of ($z_{p_i},x_{p_i}$), the polynomial model can be established by a least square method.

\renewcommand\arraystretch{1}
 \begin{table}[!hptb]
\centering
\caption{ The absolute relative error of 19th order polynomial model for simulating non-normal distributions}
\label{results}
\begin{tabular}[l]{@{}ccccc}
\toprule
   Probability distribution~~~~~ & \multirow {1}{*}{Parameters~~}  & Probit range $[\alpha,1-\alpha]$  & Maximum value of $\varepsilon_{p_k}$(\%)  \\
\midrule
\multirow {2}{*}{Gamma~~~~~}    	&	$1\leq a\leq 100$~~	    &	\multirow {2}{*}{$[10^{-4},1-10^{-4}]$}		&	\multirow {2}{*}{$0.92$}	\\
   	&	$1\leq b\leq 100$~~	    &		&		&	\\
	&		\\

\multirow {2}{*}{Beta~~~~~} 	    &	$2\leq a\leq 20$~~  	        &	\multirow {2}{*}{$[10^{-4},1-10^{-4}]$}		&	\multirow {2}{*}{$2.3\times 10^{-4}$}	\\
                           &	$2\leq b\leq 20$~~	        &	&		&		\\
	&		\\
\multirow {2}{*}{Beta~~~~~} 	    &	$1\leq a\leq 2$~~         	&	\multirow {2}{*}{$[10^{-3},1-10^{-3}]$}	&	\multirow {2}{*} {$6\times 10^{-6}$}	\\

    &	$1\leq b\leq 2$~~                   	&		&		&		\\
	&		\\

\multirow {2}{*}{Weibull~~~~~}	&	$1\leq a\leq 100$~~       	&	\multirow {2}{*}{$[10^{-4},1-10^{-4}]$}		&	\multirow {2}{*}{$0.92$}	\\
&	$1\leq b\leq 100$~~         	&		&		&		\\
	&		\\

\multirow {2}{*}{Lognormal~~~~~}    	&	$0.01\leq a\leq 200$~~	    &	\multirow {2}{*}{$[10^{-4},1-10^{-4}]$}		&	\multirow {2}{*}{$0.060$}	\\
   	&	$0.01\leq b\leq 4$~~	    &		&		&	\\
	&		\\

\multirow {2}{*}{F~~~~~}         	&	$4\leq d_1\leq 100$~~       	&	\multirow {2}{*}{$[10^{-4},1-10^{-4}]$}		&	\multirow {2}{*}{$0.090$}	\\
        	&	$4\leq d_2\leq 100$~~       	&		&		&		\\
	&		\\
\multirow {1}{*}{Uniform~~~~~}	&	$a=0$~$b=1$                      	&	\multirow {1}{*}{$[10^{-3},1-10^{-3}]$}	&	\multirow {1}{*}{$0.0035$}	\\

\multirow {1}{*}{Gumbel~~~~~}	&	$a=0$~$b=1$                      	&	\multirow {1}{*}{$[10^{-4},1-10^{-4}]$}	&	\multirow {1}{*}{$0.013$}	\\
\multirow {1}{*}{Logistic~~~~~}	&	$a=0$~$b=1$                      	&	\multirow {1}{*}{$[10^{-4},1-10^{-4}]$}	&	\multirow {1}{*}{$6.6\times 10^{-5}$}	\\
T~~~~~	&	$1\leq \nu\leq 100$~~                       	&	$[10^{-4},1-10^{-4}]$			&	$0.064$	\\
Chi squared~~~~~	&	$2\leq k\leq 100$~~                           	&	$[10^{-4},1-10^{-4}]$		&	$0.94$	\\
Rayleigh~~~~~	&	$\sigma=1$~~                          	&	$[10^{-4},1-10^{-4}]$		&	$0.0063$	\\
Exponential~~~~~	&	$\lambda=1$~~	                                    &	$[10^{-4},1-10^{-4}]$		&	$0.95$	\\
\bottomrule
\end{tabular}
\end{table}

Here, a heuristic method is put forward to establish the polynomial by percentile matching:
\begin{description}
  \item[1.] Choose  14, 16 and 15 percentage points of $p_i$ evenly over the interval $[\alpha, 0.01)$, $[0.01, 0.99)$ and $[0.99, 1-\alpha]$ respectively, ($\alpha$ is a small value of percentage).
  \item[2.]  For each percentage $p_i$, calculate the corresponding percentiles of $F(x)$ and $\Phi(z)$ as stated in Eq.\eqref{Inverse}.
  \item[3.] Using the resulted 45 pair values of $(z_{p_i},x_{p_i})$, a 19th order polynomial can be established by a least square method.
\end{description}

To check the performance of the proposed method,  the absolute relative error between the percentile from the original distribution and the one from  polynomial model is calculated:
\begin{equation}\label{percentileerror}
\begin{split}
 &\varepsilon_{p_i}=\Bigg|\frac{x_{p_i}^*-x_{p_i}}{x_{p_i}}\Bigg|\times 100[\%], ~~~~~x_{p_i}=F^{-1}({p_i}),\\
 &x_{p_i}^{*}=\sum_{k=0}^{n}a_kz_{p_i}^k,~~z_{p_i}=\Phi^{-1}(p_i).
\end{split}
\end{equation}

Testing for various probability distributions in Matlab, such as: Gamma distribution $Gamma(a,b)$, Beta distribution  $Beta(a,b)$, F-distribution $F(d_1,d_2)$, Lognormal distribution $lnN(a,b)$, Weibull distribution $Weibull(a,b)$, Uniform distribution $U(a,b)$, Gumbel distribution $Gumble(a,b)$, Logistic distribution $Logistic(a,b)$, T-distribution $T(\nu)$, Chi squared distribution $chi2(k)$, Rayleigh distribution $Rayleigh(\sigma)$ and Exponential distribution $Exp(\lambda)$. the results are shown in Table \ref{results}.

As shown in Table \ref{results}, the tested non-normal distributions are well approximated by the 19th order polynomial model. Taking the Gamma distribution $Gamma(a,b)$ for example, the error $\varepsilon_{p_i}$ for the family of Gamma distribution with parameters $1\leq a\leq 100$, $1\leq b\leq 100$ is all less than $0.92\%$, demonstrating a satisfactory accuracy.


However, if the polynomial model is employed for data fitting , a large number of samples should be provided to estimate a precise value of the percentile $x_{p_k}$. With a small to moderate sample size, the PWM matching method is more favourable.

\section{Evaluating the correlation coefficient}
Suppose $X_1$ and $X_2$ are two correlated random variables. Let $X_1$, $X_2$ both be approximated by a polynomial of degree $n$:
 \begin{equation}
   \begin{split}
     X_1 &=a_{1,0}+a_{1,1}Z_1+\cdots+a_{1,n}Z_1^n \\
     X_2 &=a_{2,0}+a_{2,1}Z_2+\cdots+a_{2,n}Z_2^n
   \end{split}
 \end{equation}
 To ensure a desired correlation coefficient $\rho_x$ between $X_1$ and $X_2$, a suitable correlation coefficient $\rho_z$ between $Z_1$ and $Z_2$ should be determined. This problem is known as the Nataf transformation\cite{NatafDKL,Nataf2} or Normal-To-Anything (NORTA) algorithm\cite{NORTA1,NORTA2}, which requires solving the following integral equation:
\begin{equation}\label{integralequation}
 \rho_x\sigma_1\sigma_2+\mu_1\mu_2 = E[X_1X_2]= \int\int F^{-1}_{1}[\Phi(Z_1)]F^{-1}_{2}[\Phi(Z_2)]\phi(Z_1,Z_2,\rho_z) dZ_1dZ_2,
\end{equation}
where  $\mu_i$, $\sigma_i$ are the mean and standard deviation of $X_i$ ($i=1,2$), $\phi(Z_1,Z_2,\rho_z)$ is the joint PDF of two normal random variables with correlation coefficient $\rho_z$. Generally, this equation is difficult to be analytically solved. While, using the polynomial normal transformation technique, the problem can be greatly simplified.

The formulae of the product moments of two correlated standard normal variables are as follows \cite{Pearson}:
\begin{equation}\label{Product}
\begin{split}
 & E\left(Z_1^{2s}Z_2^{2t}\right) = \frac{(2s)!(2t)!}{2^{s+t}}\sum_{j=0}^{min(s,t)}\frac{(2\rho_z)^{2j}}{(s-j)!(t-j)!(2j)!} \\
 & E\left(Z_1^{2s+1}Z_2^{2t+1}\right) = \rho_z\frac{(2s+1)!(2t+1)!}{2^{s+t}}\cdot \sum_{j=0}^{min(s,t)}\frac{(2\rho_z)^{2j}}{(s-j)!(t-j)!(2j+1)!}  \\
 & E\left(Z_1^{2s}Z_2^{2t+1}\right) = E\left(Z_1^{2s+1}Z_2^{2t}\right)=0 \\
 & E\left(Z_i^{2s}\right) =\frac{(2s)!}{2^s\cdot s!}~~~~i=1,2.
  \end{split}
\end{equation}
 where $s$ and $t$ are nonnegative integers.

Then, the following derivation is performed:
\begin{equation}\label{DDD}
  \begin{split}
    & \rho_x\sigma_1\sigma_2+\mu_1\mu_2=E(X_1X_2) =E\left[\left(\sum_{i=0}^{n}a_{1,i}Z_{1}^{i}\right) \left(\sum_{j=0}^{n}a_{2,j}Z_{2}^{j}\right)    \right]\\
    &= \Big(a_{1,0}~~  a_{1,1}~  \cdots  a_{1,n}\Big)
     E\begin{pmatrix} 1    & Z_2  & \cdots & Z_2^n \\
                 Z_1             & Z_1Z_2  & \cdots & Z_1Z_2^n \\
                \vdots           & \vdots         & \cdots & \vdots        \\
                 Z_1^n           & Z_1^nZ_2  & \cdots &Z_1^nZ_2^n
  \end{pmatrix}
  \begin{pmatrix} a_{2,0} \\
                a_{2,1} \\
                \vdots      \\
                a_{2,n}
  \end{pmatrix}.
  \end{split}
\end{equation}
Substitute Eq.\eqref{Product} into Eq.\eqref{DDD}, and perform matrix multiplication, a polynomial equation in $\rho_z$ of degree $n$ can be obtained:
\begin{equation}\label{wwww}
\rho_x\sigma_1\sigma_2+\mu_1\mu_2=b_n\rho_z^n+\cdot+b_{i}\rho_z^{i}+\cdots+b_1\rho_z+b_0,
\end{equation}
where $b_i$ ($i=0,\dots,n$) are sums of product of $a_{1,k_1}$ and $a_{2,k_2}$. A 19th-order polynomial is presented in \nameref{appendix}.




Then, the correlation coefficient $\rho_x$ is expressed as a polynomial function of $\rho_z$. As indicated by Lemma 3 in reference\cite{NatafDKL}, the feasible correlation coefficient $\rho_x$ cannot take any value between $-1$ and 1, and is located in a subinterval of $[-1,1]$, i.e. $-1\leq \underline{\rho_x}\leq \rho_x \leq\overline{\rho_x}\leq 1$. Based on the derived polynomial equation, the lower and upper bounds of $\rho_x$ can be determined by setting $\rho_z=-1,1$.

Furthermore, some properties of the function relationship between $\rho_z$ and $\rho_x$ have been proved:
\begin{description}
  \item[1.] $\rho_x$ is a strictly increasing function of $\rho_z$ (Lemma 1 in reference\cite{NatafDKL}).
  \item[2.] $\rho_z\geq 0$ ($\leq 0$) implies $\rho_x\geq 0$ ($\leq 0$) (Proposition 1 in reference \cite{NORTA1}), that's to say, $\rho_z\rho_x\geq 0$ .
\end{description}
Thus, for a given correlation coefficient $\rho_x$ between $X_1$ and $X_2$, the associated $\rho_z$ can be estimated by solving the polynomial equation, and a valid solution is restricted by the following conditions:
\begin{equation}
  -1\leq\rho_z\leq1~~and~~\rho_z\rho_x\geq0
\end{equation}
As long as $X_1$ and $X_2$ can be well approximated by the polynomial model,  $\rho_z$ for a given $\rho_x$ can be estimated with a satisfactory accuracy. 



Suppose $\bm{X}=(X_1,\dots, X_i, \dots, X_m)^T$ is an $m$-dimensional random vector with correlation matrix $\bm{R_X}$, the steps for generating $\bm{X}$   are as follows:
\begin{enumerate}
\item[(1)] Approximate each random variable $X_i$ by the polynomial model, determine the coefficients by the PWM matching method or percentile matching method.
\item[(2)] Calculate $\rho_z(i,j)$ ($i\neq j$)  for each entry $\rho_x(i,j)$  in $\bm{R_X}$, obtain $\bm{R_Z}$, which is the equivalent correlation matrix in the standard normal space.
\item[(3)] Generate $m$-dimensional independent standard normal vector: $\bm{U}=(U_1,\cdots, U_i, \cdots, U_m)^T$, transform $\bm{U}$ into the correlated standard normal vector $\bm{Z}=(Z_1\cdots Z_i \cdots Z_m)^T$ with correlation matrix $\bm{R_Z}$ by performing $\bm{Z=LU}$. $\bm{L}$ is the lower triangular matrix resulted from Cholesky decomposition, that is, $\bm{R_Z=LL^T}$.
\item[(4)] Convert $Z_i$ into $X_i$ through the marginal transformation $X_i=F^{-1}_i[\Phi(Z_i)]$, the random vector $\bm{X}$ with the prescribed marginal distributions and correlation matrix $\bm{R_X}$ is obtained.
\end{enumerate}
The procedures can be expressed as follows:
\begin{equation}\label{NatafP}
 \left(
 \begin{array}{c}
               U_1  \\
               \vdots \\
               U_i\\
               \vdots\\
               U_m
             \end{array}
     \right)
    \xlongrightarrow{
    \begin{array}{c}
               _{_{\bm{R_Z=LL^T}} } \\
               _{\downarrow}\\
               _{_{\bm{Z}=\bm{L}\bm{U}}}
             \end{array}}
         \left(
 \begin{array}{c}
               Z_1  \\
               \vdots \\
               Z_i\\
               \vdots\\
               Z_m
             \end{array}
     \right)      \xlongrightarrow{X_i=F_i^{-1}(Z_i)}
         \left(
 \begin{array}{c}
               X_1  \\
               \vdots \\
               X_i\\
               \vdots\\
               X_m
             \end{array}
     \right)
\end{equation}

As long as the marginal distribution $F_i(X_i)$ can be well approximate by polynomial model, and the correlation matrix  $\bm{R_X}$  is a positive definite symmetric matrix (see Proposition 3 in reference\cite{NORTA1}), the correlated random vector $\bm{X}$ can be generated by the proposed method.

\section{Numerical example}
\subsection{Simulating non-normal distributions}
The polynomial models of degree $3$, $5$, $11$ are employed to simulate the Beta distribution $Beta(2,2)$ respectively. The coefficients are determined by PWM matching method in Section 2.1, and the coefficients of the $11$th polynomial model are shown in Table \ref{table33}.  The PDFs of the random numbers generated by polynomial models are depicted in Fig. \ref{figurebeta}.
\begin{table}[hptb]
\setlength{\abovecaptionskip}{0pt}
\centering
\caption{ The coefficients of $11$th order polynomial model$--$$Beta(2,~2)$}
\label{table33}
\begin{tabular}{p{0.5cm}p{4.5cm}p{0.5cm}p{4.5cm}}
\hline $k$  & $~~~~~~~~~~~~~~~a_k$ & $k$  & $~~~~~~~~~~~~~a_k$  \\
\hline
11	&	$-2.76217093165881\times 10^{-8}$	&	5	&	$0.00120213182254751$	\\
10	&	$-8.53583161405554\times 10^{-10}$	&	4	&	$1.41667915745929\times 10^{-7}$	\\
9	&	$1.75657290203781\times 10^{-6}$	&	3	&	$-0.0192416046002625$	\\
8	&	$1.63734319656302\times 10^{-8}$	&	2	&	$-6.4707982562276\times 10^{-8}$	\\
7	&	$-0.0000555670500494067$	&	1	&	$0.265961312977451$	\\
6	&	$-8.51019206906232\times 10^{-8}$	&	0	&	$0.500000003658777$	\\
\hline
\end{tabular}
\end{table}

\setcounter{figure}{0}
\begin{figure}[!htbp]
  \centering
  \includegraphics[width=8.89cm]{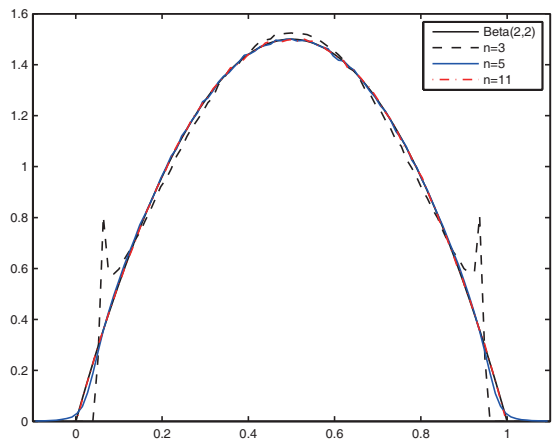}\\
  \caption{ The PDFs of Beta distribution $Beta(2,~2)$}
  \label{figurebeta}
\end{figure}

$10^4$ values of $p_i$ are evenly chosen in the probit range $[0.001, 0.999]$. Calculate $\varepsilon_{p_i}$ for each percentage $p_i$ ($i=1,\dots,10^4$) as stated in Eq.\eqref{percentileerror}, the average value,the minimum value, and  the maximum value of $\varepsilon_{p_i}$  are presented in Table \ref{table44}.
\begin{table}[hptb]
\setlength{\abovecaptionskip}{0pt}
\centering
\caption{ The values of $\varepsilon_{p_i}$ for simulating $Beta(2,2)$}
\label{table44}
\begin{tabular}{p{2.5cm}p{2.5cm}p{2.5cm}p{2.5cm}}
\hline Degree ($n$)  & Average($\%$) & Minimum($\%$) & Maximum($\%$)   \\
\hline
$3$  & $1.2$ & $1.8\times 10^{-5}$ &$389$   \\
$5$  & $0.15$ & $8.0\times 10^{-6}$ &$54$   \\
$11$  & $4.1\times 10^{-4}$ & $2.0\times 10^{-9}$     &$0.095$    \\
\hline
\end{tabular}
\end{table}

As shown in Fig. \ref{figurebeta} and Table \ref{table44}, due to the ability of controlling higher order moments, a polynomial model of higher degree provides a more accurate approximation, the PDF given by the $11$th order polynomial model is in close proximity with Beta distribution.

The TPNT model cannot provide a good approximation for the Lognormal distribution $lnN(0,1)$. Here, the polynomial models of degree 5 and 11 are employed to simulate $lnN(0,1)$. $17$ percentage points $p_i$ are evenly chosen in the interval $[0.001, 0.999]$. Estimate $z_{p_i}$ and $x_{p_i}$ as stated in Eq.\eqref{Inverse}. With the resulted 17 pair values of ($z_{p_i},x_{p_i}$), the coefficients of the polynomial model are determined by the least square method. The coefficients of the $11$th polynomial model are shown in Table \ref{table55}.

\begin{table}[hptb]
\setlength{\abovecaptionskip}{0pt}
\centering
\caption{ The coefficients of $11$th order polynomial model$--$$lnN(0,~1)$}
\label{table55}
\begin{tabular}{p{0.5cm}p{4.5cm}p{0.5cm}p{4.5cm}}
\hline $k$  & $~~~~~~~~~~~~~~~a_k$ & $k$  & $~~~~~~~~~~~~~a_k$  \\
\hline
$11$	&	$2.70587705587477\times10^{-8}$	&	$5$	&	$0.00833335056026873$	\\
$10$	&	$3.07095310533251\times10^{-7}$	&	$4$	&	$0.0416665771671574$	\\
$9$	&	$2.75245190929274\times10^{-6}$	&	$3$	&	$0.166666657138832$	\\
$8$	&	$2.46853662677270\times10^{-5}$	&	$2$	&	$0.500000016464362$	\\
$7$	&	$0.000198405765895619$	&	$1$	&	$1.00000000118744$	\\
$6$	&	$0.00138905069923553$	&	$0$	&	$0.999999999545197$	\\
\hline
\end{tabular}
\end{table}

\begin{figure}[!htbp]
  \centering
  \includegraphics[width=7.89cm]{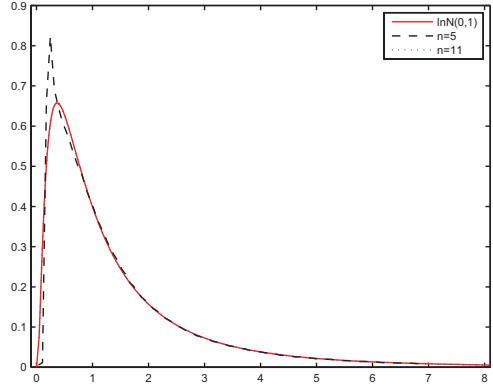}\\
  \caption{ The PDFs of Lognormal distribution $lnN(0,~1)$}
  \label{figure22}
\end{figure}

Fig. \ref{figure22} shows the PDFs of the Lognormal distribution and its polynomial model analogs. Calculate $\varepsilon_{p_i}$ as in the former case, the results are summarized in Table \ref{table66}. As can be seen, the 11th polynomial model gives better performance than the 5th one.

\begin{table}[hptb]
\setlength{\abovecaptionskip}{0pt}
\centering
\caption{The vaules of $\varepsilon_p$ for simulating $lnN(0,1)$}
\label{table66}
\begin{tabular}{p{2.5cm}p{2.5cm}p{2.5cm}p{2.5cm}}
\hline Degree ($n$)  & Average($\%$) & Minimum($\%$) & Maximum($\%$)   \\
\hline
$5$  & $3.6$ & $7.2\times 10^{-5}$ &$167$   \\
$11$  & $4.6\times 10^{-4}$ & $0$     &$0.087$    \\
\hline
\end{tabular}
\end{table}




\subsection{Generating correlated random vector}
Suppose $X_1$ and $X_2$ both follow the Lognormal distribution $lnN(0,1)$. In this case, the function relationship of $\rho_x$ and $\rho_z$ is as follows \cite{Bilognormal}:
\begin{equation}\label{BLognormal}
  \rho_x=\frac{exp(\rho_z)-1}{e-1}
\end{equation}
For a given $\rho_x$, $\rho_z$ resulted from Eq.\eqref{BLognormal} is regarded as benchmark.

\begin{figure}[!htbp]
  \centering
  \includegraphics[width=7.89cm]{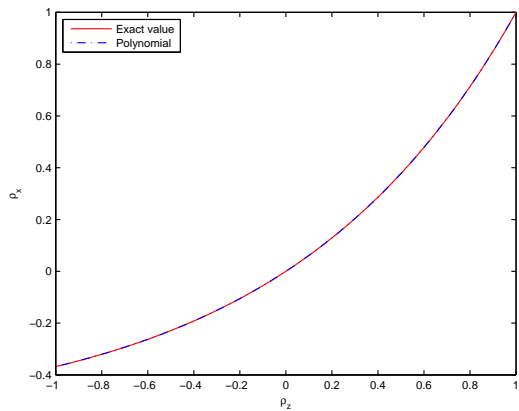}\\
  \caption{ The function relationship between $\rho_z$ and $\rho_x$}
  \label{lnNNNNN}
\end{figure}

The foregoing $11$th order polynomial model is used to simulate $X_1$ and $X_2$, of which the coefficients are presented in Table \ref{table55}. In Appendix, the coefficients $a_k$ ($k=12,\dots,19$) are set to be 0. The results are summarized in Table \ref{xiang}. The function curves are shown in Fig. \ref{lnNNNNN}.
\begin{table}[hptb]
\setlength{\abovecaptionskip}{0pt}
\centering
\caption{The correlation coefficients $\rho_z$ between $lnN(0,1)$ and $lnN(0,1)$}
\label{xiang}
\begin{tabular}{p{2.5cm}p{5.5cm}p{5.5cm}}
\hline
$\rho_x$  & $\rho_z$(Exact value) & $\rho_z$(Equation)    \\
\hline
$-0.3$  & $~~~~~-0.725~~~~~$      & $~~~~~~-0.725~~~~~~$   \\
$-0.1$  & $~~~~~-0.189~~~~~$      & $~~~~~~-0.189~~~~~~$    \\
$0.1$   & $~~~~~~0.159~~~~~$      & $~~~~~~~0.159~~~~~~$   \\
$0.3$   & $~~~~~~0.416~~~~~$      & $~~~~~~~0.416~~~~~~$    \\
$0.5$   & $~~~~~~0.620~~~~~$      & $~~~~~~~0.620~~~~~~$    \\
$0.7$   & $~~~~~~0.790~~~~~$      & $~~~~~~~0.790~~~~~~$    \\
$0.9$   & $~~~~~~0.935~~~~~$      & $~~~~~~~0.935~~~~~~$  \\
\hline
\end{tabular}
\end{table}

From Table \ref{xiang} and Fig. \ref{lnNNNNN}, it can be seen that the proposed method yields results with satisfactory accuracy.

Finally, an example of generating correlated random vector is worked. Suppose $\bm{X}=(X_1,X_2,X_3)^T$ is a three dimensional random vector, $X_1$, $X_2$, $X_3$ follow the standard normal distribution $N(0,1)$, Beta distribution $Beta(2,2)$ and Lognormal distribution $lnN(0,1)$, respectively. The desired correlation matrix $\bm{R_X}$ is:
\begin{equation}
 \nonumber \bm{R_X}
  =\begin{pmatrix}      1      &0.9    &0.5   \\
                       0.9     &1      &0.3  \\
                       0.5     &0.3    &1
  \end{pmatrix}
\end{equation}

For the normal distribution with mean $\mu$ and standard deviation $\sigma$, the coefficients of the polynomial can be regarded as: $a_0=\mu$, $a_1=\sigma$ and $a_i=0$ ($i=2,\dots,n$). The Beta distribution and Lognormal distribution are both simulated by the 11th order polynomial model,  the coefficients are tabulated in Table \ref{table33} and Table \ref{table55} respectively. The equivalent correlation matrix $\bm{R_Z}$ in the normal space is:
\begin{equation}
 \nonumber \bm{R_Z}
  =\begin{pmatrix}      1        &0.907    &0.655   \\
                       0.907     &1        &0.400  \\
                       0.655     &0.400    &1
  \end{pmatrix}
\end{equation}
$10^6$ three-dimensional random vectors are generated as stated in Eq.\eqref{NatafP}. The correlation matrix of samples is:
\begin{equation}
 \nonumber \bm{R_X^*}
  =\begin{pmatrix}      1      &0.900    &0.499   \\
                       0.900     &1      &0.300  \\
                       0.499     &0.300    &1
  \end{pmatrix}
\end{equation}
Comparing $\bm{R_X^*}$ with $\bm{R_X}$, it makes evident that the correlation matrix of samples is in close proximity to the desired one. Multivariate random variables with prescribed marginal distributions and correlation matrix are generated.
\section{Conclusions}
A polynomial normal transformation model is developed in this paper. The PWM matching method and the percentile matching method are introduced to determine the coefficients, whereby the polynomial normal transformation model is extended to a higher degree. Comparing to the extant TPNT and FPNT models, the proposed model can preserve more statistical information of the target distribution and gives more accurate approximation.

For two correlated random variables, the polynomial equations for estimating $\rho_z$ are also derived. It is shown that the original correlation coefficient $\rho_x$ can be expressed as a polynomial function of $\rho_z$. For a given $\rho_x$, the associated $\rho_z$ is estimated by solving the polynomial equation. The numerical examples demonstrate that non-normal distributions can be well approximated by the polynomial model, and random vector with prescribed correlation matrix can be generated by the proposed method.
\newpage
\section*{Appendix}\label{appendix}
\begin{equation}
\begin{split}
b_{19}=&121645100408832000a_{1,19}a_{2,19}\\
b_{18}=&6402373705728000a_{1,18}a_{2,18}\\
b_{17}=&355687428096000(a_{1,17}+171a_{1,19})(a_{2,17}+171a_{2,19})\\
b_{16}=&20922789888000(a_{1,16}+153a_{1,18})(a_{2,16}+153a_{2,18})\\
b_{15}=&1307674368000(a_{1,15}+136a_{1,17}+11628a_{1,19})(a_{2,15}+136a_{2,17}+11628a_{2,19})\\
b_{14}=&87178291200(a_{1,14}+120a_{1,16}+9180a_{1,18})(a_{2,14}+120a_{2,16}+9180a_{2,18})\\
b_{13}=&6227020800(a_{1,13}+105a_{1,15}+7140a_{1,17}+406980a_{1,19})\\
&(a_{2,13}+105a_{2,15}+7140a_{2,17}+406980a_{2,19})\\
b_{12}=&479001600(a_{1,12}+91a_{1,14}+5460a_{1,16}+278460a_{1,18})\\
&(a_{2,12}+91a_{2,14}+5460a_{2,16}+278460a_{2,18})\\
b_{11}=&39916800(a_{1,11}+78a_{1,13}+4095a_{1,15}+185640a_{1,17}+7936110a_{1,19})\\
&(a_{2,11}+78a_{2,13}+4095a_{2,15}+185640a_{2,17}+7936110a_{2,19})\\
b_{10}=&3628800(a_{1,10}+66a_{1,12}+3003a_{1,14}+120120a_{1,16}+4594590a_{1,18})\\
&(a_{2,10}+66a_{2,12}+3003a_{2,14}+120120a_{2,16}+4594590a_{2,18})\\
b_{9}=&362880(a_{1,9}+55a_{1,11}+2145a_{1,13}+75075a_{1,15}+2552550a_{1,17}+87297210a_{1,19})\\
&(a_{2,9}+55a_{2,11}+2145a_{2,13}+75075a_{2,15}+2552550a_{2,17}+87297210a_{2,19})\\
b_{8}=&40320(a_{1,8}+45a_{1,10}+1485a_{1,12}+45045a_{1,14}+1351350a_{1,16}+41351310a_{1,18})\\
&(a_{2,8}+45a_{2,10}+1485a_{2,12}+45045a_{2,14}+1351350a_{2,16}+41351310a_{2,18})\\
b_{7}=&5040(a_{1,7}+36a_{1,9}+990a_{1,11}+25740a_{1,13}+675675a_{1,15}+18378360a_{1,17}+523783260a_{1,19})\\
&(a_{2,7}+36a_{2,9}+990a_{2,11}+25740a_{2,13}+675675a_{2,15}+18378360a_{2,17}+523783260a_{2,19})\\
b_{6}=&720(a_{1,6}+28a_{1,8}+630a_{1,10}+13860a_{1,12}+315315a_{1,14}+7567560a_{1,16}+192972780a_{1,18})\\
&(a_{2,6}+28a_{2,8}+630a_{2,10}+13860a_{2,12}+315315a_{2,14}+7567560a_{2,16}+192972780a_{2,18})\\
b_{5}=&120(a_{1,5}+21a_{1,7}+378a_{1,9}+6930a_{1,11}+135135a_{1,13}+2837835a_{1,15}+64324260a_{1,17}+1571349780a_{1,19})\\
&(a_{2,5}+21a_{2,7}+378a_{2,9}+6930a_{2,11}+135135a_{2,13}+2837835a_{2,15}+64324260a_{2,17}+1571349780a_{2,19})\\
b_{4}=&24(a_{1,4}+15a_{1,6}+210a_{1,8}+3150a_{1,10}+51975a_{1,12}+945945a_{1,14}+18918900a_{1,16}+413513100a_{1,18})\\
&(a_{2,4}+15a_{2,6}+210a_{2,8}+3150a_{2,10}+51975a_{2,12}+945945a_{2,14}+18918900a_{2,16}+413513100a_{2,18})\\
b_{3}=&6(a_{1,3}+10a_{1,5}+105a_{1,7}+1260a_{1,9}+17325a_{1,11}+270270a_{1,13}+4729725a_{1,15}+91891800a_{1,17}+\\
&1964187225a_{1,19})(a_{2,3}+10a_{2,5}+105a_{2,7}+1260a_{2,9}+17325a_{2,11}+270270a_{2,13}+4729725a_{2,15}+\\
&91891800a_{2,17}+1964187225a_{2,19})\\
b_{2}=&2(a_{1,2}+6a_{1,4}+45a_{1,6}+420a_{1,8}+4725a_{1,10}+62370a_{1,12}+945945a_{1,14}+16216200a_{1,16}+\\
&310134825a_{1,18})(a_{2,2}+6a_{2,4}+45a_{2,6}+420a_{2,8}+4725a_{2,10}+62370a_{2,12}+945945a_{2,14}+\\
&16216200a_{2,16}+310134825a_{2,18})\\
b_{1}=&(a_{1,1}+3a_{1,3}+15a_{1,5}+105a_{1,7}+945a_{1,9}+10395a_{1,11}+135135a_{1,13}+2027025a_{1,15}+\\
&34459425a_{1,17}+654729075a_{1,19})(a_{2,1}+3a_{2,3}+15a_{2,5}+105a_{2,7}+945a_{2,9}+10395a_{2,11}+\\
&135135a_{2,13}+2027025a_{2,15}+34459425a_{2,17}+654729075a_{2,19})\\
b_{0}=&(a_{1,0}+a_{1,2}+3a_{1,4}+15a_{1,6}+105a_{1,8}+945a_{1,10}+10395a_{1,12}+135135a_{1,14}+2027025a_{1,16}+\\
&34459425a_{1,18})(a_{2,0}+a_{2,2}+3a_{2,4}+15a_{2,6}+105a_{2,8}+945a_{2,10}+10395a_{2,12}+135135a_{2,14}+\\
&2027025a_{2,16}+34459425a_{2,18})\\
\end{split}
\end{equation}



%
%

\end{document}